\documentstyle[11pt,epsfig]{article}
\title{\bf Multi-dimensional classical and quantum cosmology: Exact solutions, signature
transition and stabilization}
\author{S. Jalalzadeh$^1$\thanks{email: s-jalalzadeh@cc.sbu.ac.ir}
, F. Ahmadi$^1$\thanks{email: fa-ahmadi@cc.sbu.ac.ir} and H. R.
Sepangi$^{1,2}$\thanks{email: hr-sepangi@cc.sbu.ac.ir}
\\ $^1${\small Department of Physics, Shahid Beheshti University, Evin, Tehran 19839, Iran}\\$^2${\small
Computational Physical Sciences Research Laboratory,  Department
of Nano-science,}\\ {\small Institute for Studies in Theoretical
Physics and Mathematics,}\\ {\small P.O. Box 19395-5746, Tehran,
Iran }}
\begin{document}
\maketitle 

\begin{abstract}
We study the classical and quantum cosmology of a
$(4+d)$-dimensional spacetime  minimally coupled to a scalar field
and present exact solutions for the resulting field equations for
the case where the universe is spatially flat. These solutions
exhibit signature transition from a Euclidean to a Lorentzian
domain and lead to stabilization of the internal space, in
contrast to the solutions which do not undergo signature
transition.  The corresponding quantum cosmology is described by
the Wheeler-DeWitt equation which has exact solutions in the
mini-superspace, resulting in wavefunctions peaking around the
classical paths. Such solutions admit parametrizations
corresponding to metric solutions  of the field equations that
admit signature transition.  \vspace{0.5cm}
\end{abstract}\pagebreak
\tableofcontents
\section{Introduction}
The interest in multi-dimensional cosmology has its roots in the
Kaluza-Klein (KK) idea of geometric unification of interactions
and has been a source of inspiration for numerous authors over its
long history \cite{bleyer1}. Traditionally, the extra dimension in
KK theories are assumed to be periodic and can be made arbitrarily
small (of the order of the Plank length), being curled up into a
closed topology and hence undetectable. However, the relaxation of
this condition has caused interesting physics to appear over the
past few years in the form of theories which may be categorized as
having large extra dimensions. A flood of papers have appeared to
address various consequences of having large extra dimensions,
motivated by the work of Randall-Sundrum \cite{randall} and
Arkani-Hamed {\it et.al.} \cite{arkanihamed}. In the former, the
authors investigated a string theory inspired $5D$ Anti-de-Sitter
(AdS) bulk where our universe appears as an embedded $4D$ singular
hypersurface or a 3-brane. In the latter however, a
multi-dimensional theory is considered as the fundamental theory
in which the scale of gravity is taken to be the gauge unification
scale rather than  $M_{pl}$ in $4+d$ dimensions where $d$
represents the number of extra dimensions.  A notable achievement
of these investigations has been to offer a possible explanation
for the resolution of the hierarchy problem; the huge disparity in
size between various fundamental constants in nature. Yet another
approach to models with a large extra dimension comes from the
Space-Time-Matter (STM) theory where our $4D$ world is embedded in
a $5D$ manifold devoid of matter \cite{wesson}. In this theory,
the matter in $4D$ results from purely geometrical considerations
in $5D$, providing us with a theory offering a possible
explanation of the origin of matter in the universe.  The question
of the detection of these extra dimension will have to be
addressed in the future, perhaps in an indirect way using high
energy particle accelerators.

The class of problems dealing with  higher  dimensional  cosmology
with a compactified extra dimension have been used by some authors
to address issues like signature transition in classical and
quantum cosmology \cite{darabi}.  It would then seem natural to
investigate the same issues in theories with large extra
dimensions. Naturally, the problem of the stabilization of these
extra dimensions and its relation to signature transition would
also be the questions  worth investigating. This is in contrast to
the Randall-Sundrum brane-world scenarios where issues like
signature transition and quantum cosmology are fraught with
subtleties. For example, in the brane-world scenarios, signature
transition has been shown to be incompatible with the $Z_2$
symmetry, an important, though not a necessary feature of the
theory. Also, in studying the quantum cosmology of such models the
question arises as to should one consider the bulk first and the
creation of the brane later or should they both come into being at
the same time. For a more detailed discussion of the above issues,
the reader is referred to \cite{coule}.

In a multi-dimensional cosmology, a question of interest is that
of the stabilization of the extra dimensions and various methods
have been employed to address this issue. For example, in
\cite{kriskiv} stabilization is achieved by using a Casimir-like
potential, or in \cite{gunther} by using a scalar field having an
effective potential with a global minimum. The same goal has been
achieved in \cite{wald} via a balance between vacuum energy, guage
fields and the curvature of the internal space.

In this paper, we have considered  a multi-dimensional cosmology
minimally coupled to a scalar field, described by a potential
possessing a global minimum. This feature turns out to be
essential in the study of signature transition and stabilization
in our model. The presence of a scalar field in the action in the
bulk space is generally required for the dynamical stabilization
of the theory \cite{mbelek,gunther}. Also, the scalar field plays
the double role of providing an effective mass to the radion field
during inflation and acting as an inflaton which essentially
reheats the Universe before nucleosynthesis. Such a potential has
previously been used in the context of a $4D$
Friedmann-Robertson-Walker (FRW) cosmology where a
self-interacting scalar field is coupled to Einstein's field
equations with a potential containing a Sinh-Gordon scalar
interaction \cite{dereli}. For the case of a spatially flat
universe, these equations are then solved exactly for the scalar
field and the scale factor as dynamical variables, giving rise to
cosmological solutions with a degenerate metric, describing a
continuous signature transition from a Euclidean domain to a
Lorentzian spacetime. The case for a non-flat universe is
addressed in \cite{ghafoori}.  The scalar field potential used in
the present work is that given in \cite{dereli}.  As it turns out,
the scalar field plays another  role, namely, limiting the number
of the internal degrees of freedom.

The paper is organized as follows: In section two the salient
features of the model is presented and  contact is made to address
the hierarchy problem. Section three is devoted to the solution of
the field equations and the resulting classical cosmology with a
discussion of signature transition. Section four deals with the
question of stabilization while in section five the quantum
cosmology of the model is studied by presenting solutions to
Wheeler-Dewitt (WD) equation. In the last section conclusions are
drawn.
\section{The Model}

Let us start by considering a cosmological model in which the
spacetime is assumed to be of Freedmann-Robertson-Walker type with
a $d$-dimensional internal space. We adopt a chart $\{\beta, x^i,
x^\mu\}$ where $\beta, x^i$  and $x^\mu$ $(i = 1 ,  2 , 3; \mu =
4, \cdots, d + 4 )$ represent the laps function, the external
space coordinates and the internal space coordinates respectively.
The metric is given by
\begin{equation}
g = -\beta d \beta \otimes d \beta + \frac{{\bar{R}}^2 ( \beta )
}{[ 1 + (\frac{k}{4}) r^2 ]^2 } d x^i \otimes d x^i + {\bar{a}}^2
( \beta ) g ^{(d) } \label{eq1}
\end{equation}
defined on a bulk manifold with warped product topology
\begin{equation}
M = M_{ 3 + 1 } \times M_d \nonumber
\end{equation}
with $ D = 4 + d $ being the total number of dimensions,  $ k = 1
, 0 , -1 $ representing the usual spatial curvature,
$\bar{R}(\beta)$ and $ \bar{a}(\beta)$ are  the scale factor of
the universe and the radius of the $d$-dimensional internal space
respectively and $g^{(d)}$ is the metric for the internal space,
assumed to be Ricci-flat. We write the action functional as
\begin{equation}
{\cal S} = \frac{1}{2{\kappa_D}^2}\int_M d^D x \sqrt{-g} {\cal R}[
g ] -\frac{1}{2} \int_M d^D x \sqrt{-g} \left[ - \left(
\frac{\partial{\tilde{\phi}}}{\partial{\beta}} \right)^2 + 2U (
\tilde{\phi} ) \right] + { \cal S }_{YGH} \label{eq2}
\end{equation}
where $\kappa_D$ is the $D$-dimensional gravitational constant,
$\tilde{\phi}$ is a minimally coupled homogeneous scalar field
with the potential $U(\tilde{\phi})$ and ${ \cal S }_{YGH}$ is the
usual York-Gibbons-Hawking  boundary term \cite{gibbons}. The
signature of the metric (\ref{eq1}) is Lorentzian for $ \beta
> 0 $ and Euclidean for $ \beta < 0 $. For positive values of $
\beta $ one can recover the cosmic time by setting $ t =
\frac{2}{3} { \beta }^{\frac{3}{2}}$ and the corresponding metric
becomes
\begin{equation}
g = - d t \otimes d t + \frac{R^2 ( t ) }{[ 1 + (\frac{k}{4}) r^2
]^2 } d x^i \otimes d x^i + a^2 ( t ) g ^{ ( d ) } \label{eq3}
\end{equation}
where $ R ( t ) =\bar{R} ( \beta ( t )) $ and $ a ( t ) = \bar{a}
( \beta ( t )) $ in the $\{ t , x^i , x^{ \mu }\}$ chart. We solve
our differential equations in a region that dose not include $
\beta = 0 $ and seek real solutions for $ R $ and $ a $ passing
smoothly through  the $\beta=0$ hypersurface. The Ricci scalar
corresponding to metric (\ref{eq3}) is
\begin{equation} {\cal R } =
6\left[\frac{\ddot{R}}{R} + \frac{k + {\dot{R}}^2}{R^2} \right] +
2 d\frac{\ddot{a}}{a} + d ( d - 1 )\left( \frac{\dot{a}}{a}
\right)^2 + 6 d \frac{\dot{a} \dot{R}}{a R} \label{eq4}
\end{equation}
where a dot represents differentiation with respect to t. Let $
a_0$ be the compactification scale of the internal space at the
present time and
\begin{eqnarray}
\upsilon_d \equiv \upsilon_0 \times \upsilon_i \equiv {a_0}^d
\times\int_{M_d} d^d x\sqrt{-g ^{( d )}} \nonumber
\end{eqnarray}
the corresponding total volume of the internal space. After
dimensional reduction, action (\ref{eq2}) becomes
\begin{eqnarray}
{\cal S} = - \upsilon_3 \int dt \left\{ 6{ \dot{R} }^2 \Phi R +
6\dot{R} \dot{\Phi} R^2 + \frac{d - 1}{d}
\frac{{\dot{\Phi}}^2}{\Phi} R^3 - 6 k \Phi R + {\kappa_0}^2 \Phi
R^3 \left( 2 V ( \phi ) - {\dot{\phi}}^2\right )\right \}
\label{eq5}
\end{eqnarray}
where
\begin{eqnarray}
\Phi = \frac{1}{2 {\kappa_0}^2} \left( \frac{a}{a_0} \right )^d
\nonumber
\end{eqnarray}
and we have redefined the scalar field $ \tilde{\phi} $ and its
associated potential as follows
\begin{eqnarray}
\phi = \sqrt{\upsilon_d}\tilde{\phi},  \hspace{.6 cm} V ( \phi ) =
\upsilon_d U ( \phi / \sqrt{\upsilon_d} )\nonumber
\end{eqnarray}
with $ \upsilon_3 $ being the volume of the external spatial
space.

At this point, it is appropriate to make contact with the
hierarchy problem, mentioned in the introduction \cite{arkani}. In
action (\ref{eq5}), $ \kappa_0 $ is the 4-dimensional
gravitational constant defined as
\begin{equation}
{\kappa_0 }^2= \frac{8\pi}{M^2_{Pl}} :=
\frac{\kappa^2_D}{\upsilon_d}
\end{equation}
where $ M_{Pl} = 1.22\times 10^{19}$ Gev. The conventional radius
of compactification in string theory is of order $ L_{Pl} $,
resulting in a 10-dimensional gravity scale comparable to the
4-dimensional plank scale. If we normalize $ \kappa^2_D $ in such
a way that $ \kappa^2_D = 8\pi/ M^{2 + d}_{E W} $, where $ M_{EW}
$ is the Standard Model Electroweak scale, then  $ a_0 \sim 10^{
32/d - 17}$ cm, assuming $ \upsilon_i = 1 $. This points to a
possible solution of the hierarchy problem, different from grand
unified and supersymmetric theories.

\section{Classical cosmology and exact solutions}
\begin{figure}
\centerline{\begin{tabular}{ccc}
\epsfig{figure=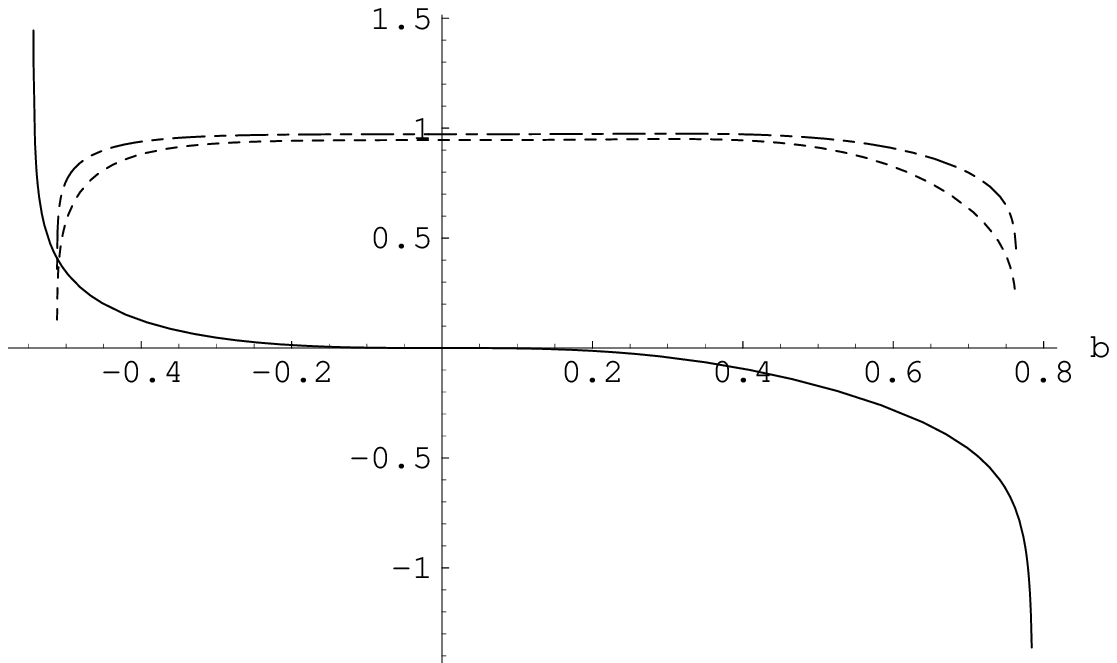,width=7cm}
 &\hspace{2cm}&
\epsfig{figure=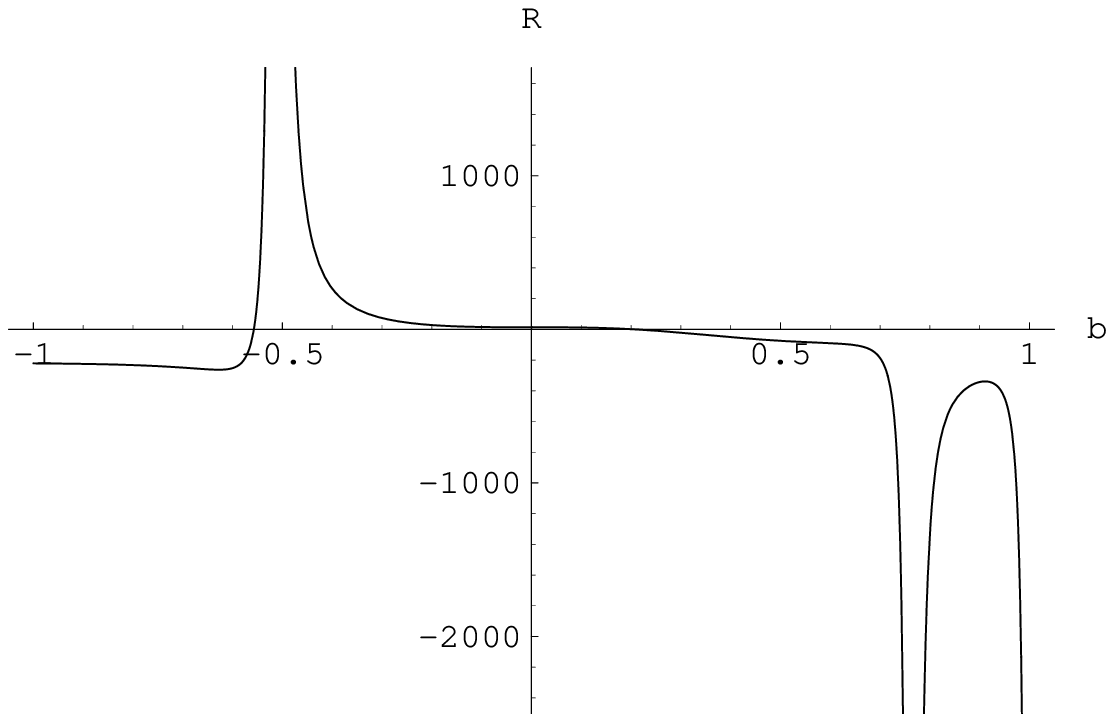,width=7cm}
\end{tabular}  }
\caption{\footnotesize Left, the scalar field (solid line), the
scale factor $R$ (dashed line) and the internal scale factor $a$
(large-small dashed line) and right, the curvature scalar $\cal R$
versus $\beta$ for $\kappa_0=1$, $d=2$, $a_1=0$, $a_2=3$, $b=1$
and $\delta=-1$.} \label{fig1}
\end{figure}

From action (\ref{eq5}), the effective Lagrangian becomes
\begin{eqnarray}
{\cal L } = -6\left\{ { \dot{R} }^2 \Phi R + \dot{R} \dot{\Phi}
R^2 + \frac{d - 1}{6 d} \frac{{\dot{\Phi}}^2}{\Phi} R^3 -  k \Phi
R + \frac{1}{3}{\kappa_0}^2 \Phi R^3 \left(  V ( \phi ) -
\frac{\dot{\phi}^2}{2}\right )\right \} \label{eq6}
\end{eqnarray}
where we have set $ \upsilon_3 = 1 $. The variation of the above
Lagrangian yields the Einstein field equations and the equation of
motion of the scalar field
\begin{eqnarray}
2\frac{\dot{\Phi}\dot{R} }{\Phi R} + \left(
\frac{\dot{R}}{R}\right)^2 + 2\frac{\ddot{R}}{R} +
\frac{\ddot{\Phi}}{\Phi} + \frac{k}{R^2} -\frac{d - 1}{2 d}\left(
\frac{\dot{\Phi}}{\Phi}\right)^2&=&{\kappa_0}^2 \left( V -
\frac{1}{2}\dot{\phi}^2 \right) \label{eq7}\\
3\left[\left(\frac{\dot{R}}{R}\right)^2 + \frac{\ddot{R}}{R} +
\frac{k}{R^2} \right] +\frac{ d -1 }{2 d} \left[ 2
\frac{\ddot{\Phi}}{\Phi} - \left( \frac{\dot{\Phi}}{\Phi}
\right)^2 + 6\frac{\dot{\Phi}\dot{R}}{\Phi R } \right]&=&
{\kappa_0}^2\left( V - \frac{1}{2} \dot{\phi}^2 \right)
\label{eq8}\\
3\left[\left(\frac{\dot{R}}{R}\right)^2 +
\frac{\dot{\Phi}\dot{R}}{\Phi R} + \frac{k}{R^2}\right] + \frac{d
- 1}{2 d}\left(\frac{\dot{\Phi}}{\Phi}\right)^2&=&{\kappa_0 }^2
\left( V + \frac{1}{2} \dot{\phi}^2 \right ) \label{eq9}\\
\ddot{\phi} + 3\frac{ \dot{R}\dot{\phi}}{R} +
\frac{\dot{\Phi}\dot{\phi}}{\Phi} +\frac{ \partial{ V } }{
\partial{ \phi}}&=&0 . \label{eq10}
\end{eqnarray}
To make the Lagrangian (\ref{eq6}) manageable, consider the the
following change of variable
\begin{eqnarray}
\Phi R^3 = mu^2 + m'v^2 + 2nuv\nonumber
\end{eqnarray}
where $ m $ , $ m' $ and $ n $ are constants and $ R = R( u , v )
, \Phi = \Phi( u , v ) $ are functions of new variables $ u ,v $.
For $ mm'-n^2 \neq{ 0 } $ , we assume
\begin{eqnarray}
\left \{ \begin{array}{l}\vspace{.2 cm}
  {\Phi}^{\rho_{+}} R^{ \sigma_{-}} = c_1 u + c_2 v \\
   {\Phi}^{\rho_{-}} R^{ \sigma_{+}}= c_3 u + c_4 v
\end{array}
\right.\label{eqs0}
\end{eqnarray}
where $ c_1 = p c_4 $ , $ c_3 = m /p c_4 $ and $ c_2 = m' /c_4 $
such that $ p = n/m' \pm \sqrt{(\frac{n}{m'})^2 - m/m'} $ and
\begin{eqnarray}
\left \{ \begin{array}{l}\vspace{.2 cm}
  \sigma_{\pm} = \frac{1}{2}\left( 3 \mp \frac{1}{\sqrt{\frac{d + 2}{3
  d}}}\right)\\
  \rho_{\pm} = \frac{1}{2} \pm \frac{3}{4}\sqrt{\frac{d + 2 }{3
  d}} \mp \frac{1}{4\sqrt{\frac{ d + 2}{3 d} }}.
\end{array}
\right.
\end{eqnarray}
Using the above transformations and concentrating on $k=0$, the
Lagrangian becomes
\begin{eqnarray}
{\cal L}=-4\left(\frac{d +2}{d +3}\right)\left \{ m\dot{u}^2 +
m'\dot{v}^2 + 2n\dot{u}\dot{v} +
\frac{1}{2}{\kappa_0}^2\left(\frac{d + 3}{d + 2}\right)( mu^2 +
m'v^2 +2nuv )(V-\frac{1}{2}\dot{\phi}^2)\right\}. \label{eq11}
\end{eqnarray}
Let us define $ \zeta $ to be the following  vector
\begin{eqnarray}
\zeta=\left (\!\! \begin{array}{l}
  u\cosh( \gamma \phi ) \\
  v\cosh( \gamma \phi ) \\
  u\sinh( \gamma \phi ) \\
v\sinh( \gamma \phi )
\end{array}\!\!\right).\label{eqs1}
\end{eqnarray}
This definition will allow us to write Lagrangian (\ref{eq11}) as
\begin{eqnarray}
{\cal L} = -\frac{{\kappa_0}^2}{\gamma ^2}\left\{
\dot{\zeta}^{\top} \Gamma \dot{\zeta} + 2\gamma ^2 \zeta^{\top}
\Gamma \zeta V \right \} \label{eq12}
\end{eqnarray}
where $ \Gamma=\left(\!\!\!\begin{array}{cc}
  J & \,\,\,0 \\
  0 & -J
\end{array}\!\!\!\right) $, $ J= \left(\!\!\!\begin{array}{cc}
  m & n \\
  n & m'
\end{array}\!\!\!\right) $, $ \gamma= \frac{\kappa_0}{2} \sqrt{\frac{d + 3}{d + 2}} $
and $\top$ represents transposition. Up to this point the
cosmological model has been rather general. However,  motivated by
the desire to find suitable smooth functions $ R(t)$, $ \Phi(t)$
and $ \phi(t)$, and in particular to stabilize the internal
degrees of freedom, we have to specify a suitable potential
$V(\phi)$. As has been discussed by various authors
\cite{gunther}, stabilization of the internal space can be
achieved if the potential  has a global minimum with respect to $
\phi $.  A Potential with such properties has been used in the
past \cite{dereli} in the context of signature transition and we
therefore adapt it here for our present purpose. We require that
the potential $V( \phi )$ has natural characteristics for small
$\phi$, so that we may identify the coefficient of $ \frac{1}{2} {
\phi}^2 $ in its Taylor expansion as a positive $ M^2$ and $ V(0)$
as a $D$-dimensional cosmological constant $\Lambda $. In this
case the effective Lagrangian can be simplified if we select a
potential that satisfies
\begin{eqnarray}
2 \gamma^2 {\zeta}^{\top} \Gamma \zeta V( \phi ) = { \zeta }^{
\top} \Gamma \Sigma \zeta \label{eq13}
\end{eqnarray}
where $\Sigma =\left(\!\!\!\begin{array}{ll}
  \,\,\,a_1 & \,\,\,\, b \\
  -b & -a_2
\end{array}\!\!\!\right) $. In terms of $\phi $, (\ref{eq13}) implies
\begin{eqnarray}
V(\phi) = \frac{(a_1 - a_2)}{4\gamma^2} + \frac{( a_1 + a_2 )}{4
\gamma ^2}\cosh{2\gamma \phi} + \frac{b}{ 2\gamma^2}\sinh{ 2\gamma
\phi}.\label{eq14}
\end{eqnarray}
Inserting the physical parameters $ \Lambda = V | _{ \phi=0} =
a_1/{2 \gamma^2} $ and $ M^2 =  \partial^2 V / \partial \phi^2 |_{
\phi =0} = a_1 + a_2 $ in the potential we find
\begin{eqnarray}
V( \phi ) = \Lambda + \frac{1}{2\gamma^2} M^2 \sinh^2{\gamma \phi}
+ \frac{b}{2 \gamma^2}\sinh{ 2\gamma \phi }. \label{eq15}
\end{eqnarray}
The first two terms in (\ref{eq14} ) give rise to a Sinh-Gordon
scalar interaction. The third term is interesting since its
presence breaks the symmetry of $ V $ under $ \phi \rightarrow
-\phi $ and is directly responsible for the signature changing
properties of the solutions to be discussed below. For $ | 2 b/M^2
| < 1 $ the potential has a minimum value
\begin{equation}
\Lambda + 4 M^2 \gamma^2 \left( \sqrt{ 1-\frac{4 b^2}{M^4}} -
1\right) \hspace{.5 cm}\mbox{at} \hspace{.6 cm}\phi = -\frac{1}{2
\gamma}\tanh^{-1} {\left( \frac{2 b}{M^2}\right)}.\label{eqq21}
\end{equation}
This minimum is believed to support the stabilization of the
internal space degrees of freedom as is discussed below. Now, if
we choose normal mode basis that diagonalize $\Sigma$ and write $
\zeta = S \alpha( t ) $ with
\begin{eqnarray}
S^{-1} \Sigma S = \Sigma _d = \left(\!\! \begin{array}{ll}
  \lambda _+ & 0 \\
  0 & \lambda_-
\end{array}\!\!\!\!\right) \nonumber
\end{eqnarray}
where
\begin{eqnarray}
S= \left(\!\! \begin{array}{cc}
  \frac{b}{ \Omega _1}  & \frac{b}{ \Omega _2} \\
  \frac{ ( \lambda _+ - a_1)}{ \Omega _1} & \frac{ (\lambda _- - a_1) }{\Omega _2}
\end{array}\!\!\right ) \nonumber
\end{eqnarray}
and
\begin{eqnarray}
\lambda _{\pm} = \frac{1}{2} \left[ a_1 - a_2 \pm \sqrt{(a_1 +
a_2)^2 -4b^2} \right] \nonumber
\end{eqnarray}
\begin{eqnarray}
 \Omega _{1} = \sqrt{b^2 - (\lambda _{+} - a_1 )^2 } \nonumber \\
 \Omega _{2} = \sqrt{(\lambda _{-} - a_1 )^2 - b^2}\nonumber
\end{eqnarray}
the Lagrangian becomes
\begin{eqnarray}
{\cal L} = -\frac{{\kappa_0}^2}{\gamma^2}\left\{
\dot{\alpha}^{\top} \Gamma \dot{\alpha} + \alpha^{ \top} \Gamma
\Sigma _d \alpha \right \}\label{eq155}
\end{eqnarray}
with the general solution for the equations of motion  given by ($
mm' - n^2 \neq 0 $)
\begin{eqnarray}
\alpha = \Upsilon _ {+}(t)A + \Upsilon _{-}(t)B
\end{eqnarray}
where A  and  B  are constant $(4 \times 1) $ vectors and $
\Upsilon_{ \pm} = \mbox{diag}(e^{ \pm \omega_{+}t},e^{\pm
\omega_{+}t},
 e^{ \pm \omega_{-}t},e^{\pm \omega_{-}t})$ with
$ \omega_{ \pm } =\sqrt{\lambda _{\pm }} $. In terms of these
solutions, the constraint equation (\ref{eq9}) becomes
\begin{eqnarray}
{\cal H} =
-\frac{{\kappa_0}^2}{\gamma^2}\left\{\dot{\alpha}^{\top} \Gamma
\dot{\alpha} - \alpha^{ \top} \Gamma \Sigma _d \alpha  \right \} =
0. \label{eq16}
\end{eqnarray}
Let us concentrate on the solutions for which $\dot{\alpha}(0) =
0$, that is $A = B$. This would then imply
\begin{equation}
\alpha = \left(\!\!\begin{array}{c}
  2A_{1} \cosh( \omega _{+}t) \\
  2A_{2} \cosh( \omega _{+}t)\\
  2A_{3} \cosh( \omega _{-}t)\\
  2A_{4} \cosh( \omega _{-}t)
\end{array}\!\!\right). \label{eqs2}
\end{equation}
 Equation (\ref{eq16}) can now be written as
\begin{eqnarray}
\left ( l^2 {A^2_{2}} - {A^2_{4}} \right) \left [
m\left(\frac{A_{3}}{A_{1}}\right)^2 + 2n\frac{A_{3}}{A_{1}} + m'
\right] = 0 \label{eq17}
\end{eqnarray}
where $ l^2=\frac{\lambda_{+}}{\lambda_{-}} $. From the vanishing
of the first parentheses we have $ A_{4} =\epsilon l A_{2}$ with
$\epsilon=\pm 1 $. Upon using (\ref{eqs2}) we find
\begin{eqnarray}
\left \{
 \begin{array}{c}
 \alpha _{2} = \epsilon l \frac{A_{2}}{A_{3}} \alpha_{1} \vspace{.2 cm}\\
  \alpha_{4} = \epsilon l \frac{A_{2}}{A_{3}} \alpha_{3}
\end{array}
\right. \label{eq18}
\end{eqnarray}
where use has been made of the relation $ \zeta_1 \zeta_4 =
\zeta_3 \zeta_2 $ resulting from (\ref{eqs1}). If we choose an
overall scale by setting $ A_{1} = 1 $, then (\ref{eqs0}),
(\ref{eqs1}) and (\ref{eq18}) result in the first class of
 solutions given by
\begin{eqnarray}
\frac{R}{R_0} = \frac{a}{a_0} = \left \{ \begin{array}{cc} \left[
\frac{\cosh^2(\omega_{+}t) - l^2\cosh^2(
\omega_{-}t)}{\cosh^2(\omega_{+}t_0 ) - l^2\cosh^2
(\omega_{-}t_0)}\right]^{\frac{1}{d ( d + 3 )} }& \hspace{2.5 cm}
\lambda_{+},
\lambda_{-} > 0 \vspace{.2 cm} \\
\left[ \frac{\cos^2(\omega_{+}t) - l^2\cos^2(
\omega_{-}t)}{\cos^2(\omega_{+}t_0) - l^2\cos^2(
\omega_{-}t_0)}\right]^{\frac{1}{d ( d + 3 )} }&\hspace{2.5 cm}
\lambda_{+}, \lambda_{-} < 0
\end{array}
\right.  \label{eq19}
\end{eqnarray}
and
\begin{eqnarray}
\phi =  \frac{1}{\gamma}\tanh^{-1}\left \{
\begin{array}{cc}
\frac{ \frac{( \lambda_{+} - a_1)}{ \Omega_1} \cosh( \omega_{+}t)
+ \frac{(\lambda_{-} - a_1)}{ \Omega_2} \epsilon l \cosh(
\omega_{-}t)} {b \left[ \frac{1}{\Omega_1} \cosh(\omega_{+}t) +
\frac{ \epsilon  l }{\Omega_2}
\cosh(\omega_{-}t)\right]} & \hspace{1 cm} \lambda_{+} , \lambda_{-} > 0 \vspace{.3 cm}  \\
\frac{ \frac{( \lambda_{+} - a_1)}{ \Omega_1} \cos( \omega_{+}t) +
\frac{(\lambda_{-} - a_1)}{ \Omega_2} \epsilon l \cos(
\omega_{-}t)}{b \left[ \frac{1}{\Omega_1} \cos(\omega_{+}t)
+\frac{ \epsilon  l }{\Omega_2} \cos(\omega_{-}t)\right]} &
\hspace{1 cm} \lambda_{+} , \lambda_{-} < 0
\end{array}
\right. \label{eq20}
\end{eqnarray}
where $R_0 $ and $ a_0  $ represent the present external and
internal scale factors and $t_0$ represents the age of the
Universe. We also have
\begin{eqnarray}
( 2{\kappa_0}^2 {R_0}^d )^{d + 3} = \left[ 2 ( c_3 + \frac{c_4
A_2}{ l } )\right]^{ 2 + ( d - 3 )\sqrt{ \frac{ d + 2}{ 3 d }}
}\left[ 2 ( c_1 + \frac{c_2 A_2}{ l } )\right]^{  ( d - 3 )\sqrt{
\frac{ d + 2}{ 3 d }} - 2 } \nonumber.
\end{eqnarray}
If the second factor in (\ref{eq17}) vanishes and setting $ A_{1}
= 1 $, we obtain the second class of solutions, namely
\begin{eqnarray}
\left \{
\begin{array}{c}
\alpha_{2} = p'\alpha_{1}\vspace{.2 cm }\\
\alpha_{4} = p'\alpha_{3}
\end{array}
\right. \label{eq21}
\end{eqnarray}
where $p' = -n/m' \pm \sqrt{ (n/m')^2 - m/m'}$. The corresponding
expressions for $R$, $a$ and $\phi$ are given by
\begin{eqnarray}
\frac{R}{R_0} = \frac{a}{a_0} =\large \left \{ \begin{array}{cc}
\left[\large \frac{\cosh^2(\omega_{+}t) -
A^2_{3}\cosh^2(\omega_{-}t)} {\cosh^2(\omega_{+}t_0) -
A^2_{3}\cosh^2(\omega_{-}t_0)}\right]^
{\frac{1}{d ( d + 3 )}} &\hspace{2 cm} \lambda_{+} , \lambda_{-} > 0 \vspace{.3 cm}  \\
\left[\frac{\cos^2(\omega_{+}t) - A^2_{3}\cos^2(\omega_{-}t)}
{\cos^2(\omega_{+}t_0) -
A^2_{3}\cos^2(\omega_{-}t_0)}\right]^{\frac{1} {d ( d + 3 )}}
&\hspace{2 cm} \lambda_{+} , \lambda_{-} < 0
\end{array}
\right. \label{eq22}
\end{eqnarray}
and
\begin{eqnarray}
 \phi = \frac{1}{\gamma}\tanh^{-1} \large \left \{
\begin{array}{cc}
\frac{\frac{(\lambda_{+}-a_1)}{ \Omega_1} \cosh( \omega_{+}t) +
A_3\frac{(\lambda_{-}-a_1)}{ \Omega_2} \cosh( \omega_{-}t)}
{b\left[\frac{1}{\Omega_1} \cosh(\omega_{+}t) + \frac{1
}{\Omega_2}
\cosh(\omega_{-}t)\right]}&\hspace{1 cm}  \lambda_{+} , \lambda_{-} > 0 \vspace{.3 cm}  \\
\frac{\frac{(\lambda_{+}-a_1)}{ \Omega_1} \cos( \omega_{+}t) + A_3
\frac{(\lambda_{-}-a_1)}{ \Omega_2} \cos( \omega_{-}t)}
{b\left[\frac{1}{\Omega_1} \cos(\omega_{+}t) +
\frac{A_3}{\Omega_2} \cos(\omega_{-}t)\right]}. & \hspace{1 cm}
\lambda_{+} , \lambda_{-} < 0
\end{array}
\right. \label{eq23}
\end{eqnarray}
 We also find
\begin{eqnarray}
( 2 {\kappa_0}^2 {R_0}^d )^{ d + 3} = \left[ 2 ( c_1 + p' c_2 )
\right]^{ ( d - 3 )\sqrt{ \frac{d + 2 }{3 d }} + 2 }\left[ 2 ( c_3
+ p' c_4 ) \right]^{ ( d - 3 )\sqrt{ \frac{d + 2 }{3 d }} - 2
}.\nonumber
\end{eqnarray}
The third class of the solutions are obtained when $ n^2 - mm' $
is zero. In this case we have
\begin{equation}
\Phi R^3 = mu^2 + m'v^2 + 2nuv = ( \sqrt{m}u + \sqrt{m'}v )^2 :=
q^2
\end{equation}
hence, following the same procedure as for the previous solutions
we adapt the following transformations
\begin{eqnarray}
\left \{
 \begin{array}{c}
{ \Phi}^{\rho_+} R^{\sigma_-} = q^{2 ( \sigma_- -\delta \rho_+ )/ (3 - \delta)} \\
{ \Phi}^{\rho_-} R^{\sigma_+} = q^{2 ( \sigma_+ -\delta \rho_- )/
(3 - \delta)}
\end{array}
\right. \nonumber
\end{eqnarray}
where $ \delta $ is real and $ \delta \neq 3 $.  Let us set $
\zeta = \left(\!\! \begin{array}{c}
  q \cosh \gamma ' \phi \\
q \sinh \gamma ' \phi
\end{array}\!\! \right ) $, $ j = \left(\!\! \begin{array}{cc}
  1 & \,\,0 \\
  0 & -1
\end{array}\!\! \right )$ and for convenience  replace $\gamma$ with $\gamma'$ in
potential (\ref{eq14}). The Lagrangian then becomes
\begin{eqnarray}
{\cal L} = -\frac{{\kappa_0}^2}{{\gamma'}^2}\left \{
{\dot{\zeta}}^{\top}j \dot{\zeta} + {\zeta }^\top \Sigma _d \zeta
\right \} \label{eq24}
\end{eqnarray}
where $$ {\gamma ' }^2= \frac{\kappa_0^2 d ( 3 - \delta
)^2}{24d(1-\delta)+4(d-1)\delta^2}.$$ The general solution in the
normal mode, $\alpha = S^{-1} \zeta$, is then given by
 \begin{eqnarray}
 \alpha = \Upsilon_+ A + \Upsilon_-B \label{eq25}
 \end{eqnarray}
where, $ A $ and $ B $ are constant $ 2\times 1 $ vectors and
$\Upsilon_{\pm } =\mbox{diag} ( e^{\pm\omega_{+}t} , e^{
\pm\omega_{-}t})$. If we choose $ \dot{\alpha}(0) = 0 $, the
constraint (\ref{eq9}) becomes
\begin{eqnarray}
 { \alpha( 0 ) }^\top \Sigma_d \alpha ( 0 ) = 0
\end{eqnarray}
and we have $ A_2 = \epsilon l A_1 $. The  solutions are then
given by
\begin{eqnarray}
\left \{
\begin{array}{cc}
\alpha _2 = 2 A_1\epsilon l \cosh ( {\frac{1}{l} \cosh^{-1}
({\frac{\alpha _1}{2 A_1}})}) &\hspace{2.5 cm} \lambda _{+},
\lambda _{-} > 0 \vspace{.2 cm}  \\
\alpha _2 = 2 A_1\epsilon l \cos( {\frac{1}{l} \cos^{-1}
({\frac{\alpha _1}{2 A_1}})}) &\hspace{2.5 cm} \lambda _{+}
,\lambda _{-} < 0
\end{array}
\right. \label{eq26}
\end{eqnarray}
and
\begin{eqnarray}
\frac{R}{R_0} =  \left(\frac{a}{a_0} \right)^{-\frac{d}{\delta}}=
\left\{
\begin{array}{cc}
 \left[\frac{ \cosh^2(\omega_{+} t ) - l^2 \cosh^2( \omega_{-} t) }
 { \cosh^2(\omega_{+} t_0 ) - l^2 \cosh^2( \omega_{-} t_0)}
 \right]^{\frac{1}{ 3- \delta}}  &\hspace{2.5 cm}  \lambda _{+},
 \lambda _{-} > 0 \vspace{.3 cm} \\
 \left[ \frac{ \cos^2 (\omega_{+} t ) - l^2 \cos^2 ( \omega_{-} t) }
 { \cos^2 (\omega_{+} t_0 ) - l^2 \cos^2 ( \omega_{-} t_0)}
 \right]^{\frac{1}{ 3- \delta}}  & \hspace{2.5 cm} \lambda _{+}, \lambda _{-} < 0
\end{array}
\right. \label{eq27}
\end{eqnarray}
with
\begin{eqnarray}
\phi =\frac{1}{\gamma}\tanh^{-1}\left \{\begin{array}{cc}
\frac{\frac{{(\lambda_{+}-a_1)}}{ \Omega_1} \cosh( \omega_{+}t) +
l\epsilon \frac{(\lambda_{-}-a_1)}{ \Omega_2}
\cosh((\omega_{-}t)}{b\left[\frac{1}{\Omega_1} \cosh(\omega_{+}t )
+ \frac{\epsilon l }{\Omega_2} \cosh(\omega_{-}t)\right]}
&  \hspace{1 cm}  \lambda_{+} , \lambda_{-} > 0 \vspace{.3 cm}  \\
\frac{\frac{(\lambda_{+}-a_1)}{ \Omega_1} \cos(\omega_{+}t ) +
l\epsilon \frac{(\lambda_{-}-a_1)}{\Omega_2}\cos( \omega_{-}t )}
{b\left[\frac{1}{\Omega_1}\cos(\omega_{+}t )+ \frac{\epsilon l }
{\Omega_2} \cos(\omega_{-}t)\right]}  & \hspace{1 cm} \lambda_{+}
, \lambda_{-} < 0
\end{array}
\right. \label{eq28}
\end{eqnarray}
where
\begin{eqnarray}
{ R_0}^{\delta} = 2 {\kappa_0}^2. \label{eq228}
\end{eqnarray}
For this class of solutions we  clearly have
\begin{equation}
\frac{a}{a_0} =
\left(\frac{R}{R_0}\right)^{-\frac{\delta}{d}}\label{eq229}
\end{equation}
in agreement with that found in \cite{mohammedi}.  It is clear
that if the parameters satisfy $ V = \frac{\partial V}{\partial
\phi} = 0 $ with $ R $, $ a $ and $ \phi $ being constants, then
for $ \Sigma $, we have degenerate  zero eigenvalues. For  such
parameters, the cosmological model becomes an eternal cosmology
with a single transition from a  Euclidean to a Lorentzian domain
at $ \beta = 0 $. For positive eigenvalues of $ \Sigma $,
signature transition dose not occur since the solutions would not
be continuous at $\beta=0$. If the product of $ \lambda_+ $ and $
\lambda_-$ be less than zero, then $ l $ becomes imaginary and the
constraint equation can not be satisfied with a real solution.
With both eigenvalues negative, equations (\ref{eq19}),
(\ref{eq22}) and (\ref{eq27}) exhibit a continuous transition from
a Euclidean to a Lorentzian domain, see figure 1. As equation
(\ref{eq19}) shows, $ R /R_0 $ and $ a /a_0 $ have the same form
and the values of $ \beta $ where $ \bar{R}( \beta ) = 0 $ control
the location of the branch points of $ \bar{R}( \beta )$. In
equations (\ref{eq20}) and (\ref{eq28}) if we choose $ \epsilon =
-1 $ and set $ \Lambda = 0$, that is, $a_1 = 0 $ then the scalar
field vanishes at the transition. The singular behavior of the
scalar field at the birth and death of the universe is reflected
in the behavior of the scalar curvature ${\cal R }$ given by
equation (\ref{eq4}). As we have shown above, we see from
equations (\ref{eq22}) and (\ref{eq23}) that the values of $A_3$
control the locations of the branch points of the the relevant
fields, otherwise, the general behavior of the solutions is the
same as that discussed above. From equation (\ref{eq228}) one sees
that due to the size of $R_0$, $\delta $ must have negative values
and this  makes the solutions finite. Figure \ref{fig1} shows the
plot of the scalar field, the scale factor $R$, the internal scale
factor $a$ and the curvature scalar $\cal R$, using equations
(\ref{eq27}) and (\ref{eq28}) for some specific values of the
parameters. The smooth transition from the negative to positive
values of $\beta$ corresponding to signature transition from a
Euclidean to a Lorentzian domain is clearly seen.

The solutions presented above also merit the following
observations. For the first and second class solutions, one
observes that in the region where $\beta<0$, one may encounter
another region over which the signs of $R^2$ and $a^2$ change
simultaneously, signifying the onset of a second signature
transition of the metric. Equations (\ref{eq19}) and (\ref{eq22})
then suggest that the exponent $d(d+3)/2$ must be an odd integer.
This in turn imposes a constraint on the value of $d$. The same
argument may be carried forward  for the third class of solutions
and again, leads to the imposition of certain limits on the values
of $d$ and $\delta$. A careful examination of the latter solutions
reveals that $d$ must be an odd integer and $\delta$ should be
represented by a negative ratio whose numerator and denominator
consist of odd integers. The above argument suggests a mechanism
through which one may put limits on the number of internal
dimensions. For example, within the context of this model, the
third class of solutions require $d=2$ in order to avoid a second
signature transition. In view of this, the present model could
also offer signature transition as a mechanism for restricting the
number of the internal degrees of freedom.
\section{Stabilization}
As was mentioned in the previous section, stabilization of the
internal space is related to the potential having a global
minimum. This issue needs to be further elaborated at this point.

To begin with, we assume that the potential has a global minimum
at zero, that is, according to equation (\ref{eqq21}) at
\begin{eqnarray}
\Lambda = 4 M^2 \gamma'^2 \left(1- \sqrt{ 1-\frac{4
b^2}{M^4}}\right). \nonumber
\end{eqnarray}
This implies a positive $\Lambda$ for the model under discussion.
Now,  from equations (\ref{eq228}) and (\ref{eq229}) we find
$$R=\Phi^{-1/\delta}.$$ Substituting this relation for $R$ into
Lagrangian (\ref{eq6}) and combining the resulting equation of
 motion for $\Phi$ and the Hamiltonian constraint, we obtain
\begin{eqnarray}
3H\dot{\Phi}+\ddot{\Phi}=\frac{\kappa_0^2\delta(\delta-3)}
{1-\delta+\delta^2\frac{d-1}{6d}} V(\phi)\Phi \label{eqq1}
\end{eqnarray}
where $H=\dot{R}/R$. Following \cite{gunther}, we now expand the
potential $V(\phi)$ about its minimum value which we have assumed
to be zero. In the zero order approximation, the right hand side
of the above equation becomes zero and a solution yields
\begin{eqnarray}
\Phi(t)=\frac{1}{2\kappa_0^2}-C_0\int_t^{t_f} dt e^{-3\int_{t_i}^t
H dt}, \label{eqq2}
\end{eqnarray}
where $C_0$ is the initial value of $\dot{\Phi}$ at $t=t_i$ where
$i$ stands for initial and $t_f$ refers to the value of $t$ close
to the region signifying the singular behavior of our solutions as
depicted in figure 1. The constant $C_0$ is chosen such that
$\Phi(t\rightarrow\infty)\rightarrow 1/2\kappa_0^2$. This choice
corresponds to the stabilization of the internal space
$a\rightarrow a_0$. That is, in the zero order approximation the
dynamical stabilization is achieved if the integral in equation
(\ref{eqq2}) converges, otherwise, decompactification would occur.
Equation (\ref{eqq2}) can be recast into a simpler form if we
substitute for $H$ and do the inner integral. The result is
\begin{eqnarray}
\Phi=\frac{1}{2\kappa_0^2}-C_0
R_0^3\int_t^{t_f}\frac{dt}{R^3}\label{eqq3}
\end{eqnarray}
where $R_i=R(t=t_i)$. An inspection of all the solutions for
$\lambda_{\pm}<0$  shows that their substitution in equation
(\ref{eqq3}) results in a convergent integral all the way close to
the region where the solutions start to diverge. However, for the
solutions corresponding to $\lambda_{\pm}>0$, the integral above
diverges and there would be no stabilization.

It would be interesting to note that the solutions corresponding
to $\lambda_{\pm}>0$ were shown in the previous section not to
undergo signature transition. One would therefore lead to the
conclusion that within the context of the present model, signature
transition and stabilization of the internal degrees of freedom
are correlated.
\section{Quantum cosmology and wave packets}
Let us now turn to the study of quantum cosmology of the model
presented above.  As was shown in the previous section, there are
three classes of solutions, the first two comprise  $mm'-n^2\ne 0$
and the third results from $mm'-n^2$ being zero. The former
classes of solutions are represented by the Lagrangian
(\ref{eq155}). An examination of this equation and the
corresponding Hamiltonian (\ref{eq16}) reveals their complicated
structure. One would therefore expect that the resulting quantum
cosmology become equally complicated and any hope of finding
analytical solution to the resulting WD equation would be in vain.
However, this is not the case for the latter class of solutions.
This motivates us to concentrate on the quantum cosmology
corresponding to the classical solutions represented by equation
(\ref{eq27}), since it can be cast into an
oscillator-ghost-oscillator system whose solutions are easily
obtained and are well know. The relevant Lagrangian is given by
equation (\ref{eq24}) and can be written as
 \begin{equation}
 {\cal L}
 =-\frac{k_0^2}{\gamma'^2}\left\{\dot{\alpha}_1^2-\dot{\alpha}_2^2
 -{\omega_{+}}^2\alpha_1^2+{\omega_{-}}^2\alpha_{2}^{2}\right\}.  \label{4.1}
\end{equation}
The Hamiltonian can then be obtained by the usual Legender
transformation where, upon quantization $p_1\rightarrow
-i\partial/\partial\alpha_1$ etc., one arrives at the WD equation
describing the  corresponding quantum cosmology
\begin{eqnarray}
{\cal H}
\Psi(\alpha_{1},\alpha_{2})=\left\{-\frac{\partial^{2}}{\partial
\alpha_{1}^{2}}+\frac{\partial^{2}}{\partial
\alpha_{2}^{2}}+{\omega_{+}}^2\alpha_{1}^{2}-{\omega_{-}}^2\alpha_{2}^{2}
\right\}\Psi(\alpha_{1},\alpha_{2})=0. \label{4.3}
\end{eqnarray}
This equation is separable in the minisuperspace variables and a
solution can be written as
\begin{equation}
\Phi_{n_{1},n_{2}}(\alpha_{1},\alpha_{2})=u_{n_{1}}(\alpha_{1})v_{n_{2}}(\alpha_{2})
\label{4.4}
\end{equation}
where
\begin{eqnarray}
u_{n}(\alpha_{1})=\left(\frac{\omega_{+}}{\pi}\right)^{1/4}\left[\frac{H_{n}\left(
\sqrt{\omega_{+}}\alpha_{1}\right)}{\sqrt{2^{n}n!}}\right]e^{
-\omega_{+}\alpha_{1}^{2}/2} \label{4.5}
\end{eqnarray}
\begin{eqnarray}
v_{n}(\alpha_{2}) =\left(
\frac{\omega_{-}}{\pi}\right)^{1/4}\left[\frac{H_{n}\left(
\sqrt{\omega_{-}}\alpha_{2}
\right)}{\sqrt{2^{n}n!}}\right]e^{-\omega_{-}\alpha_{2}^{2}/2}.
\label{4.6}
\end{eqnarray}
In these expressions $ H_{n}(x)$ is a Hermite Polynomial. The zero
energy condition, ${\cal H}=0$, then yields
\begin{eqnarray}
(n_{1}+1/2)\omega_{+}=(n_{2}+1/2)\omega_{-},  \hspace{1cm}
n_{1},n_{2}=0,1,2,\cdots.           \label{4.7}
\end{eqnarray}
The set $ \{\Phi _{n_{1},n_{2}}(\alpha_1,\alpha_2)\} $ forms a
closed span of the zero sector subspace of the Hilbert space $
\mathrm{L}^{2} $ of measurable square-integrable functions on
${\mathbf R}^{2} $ with the usual inner product defined as
\begin{eqnarray}
\int \Phi_{n_{1},n_{2}}(\alpha_{1},\alpha_{2})
\Phi_{n_{1}',n_{2}'}(\alpha_{1},\alpha_{2}) d\alpha_{1}d\alpha_{2}
=\delta_{n_{1},n_{1}^{'}}\delta_{n_{2},n_{2}^{'}}\nonumber
\end{eqnarray}
\begin{figure}
\centerline{\begin{tabular}{ccc}
\epsfig{figure=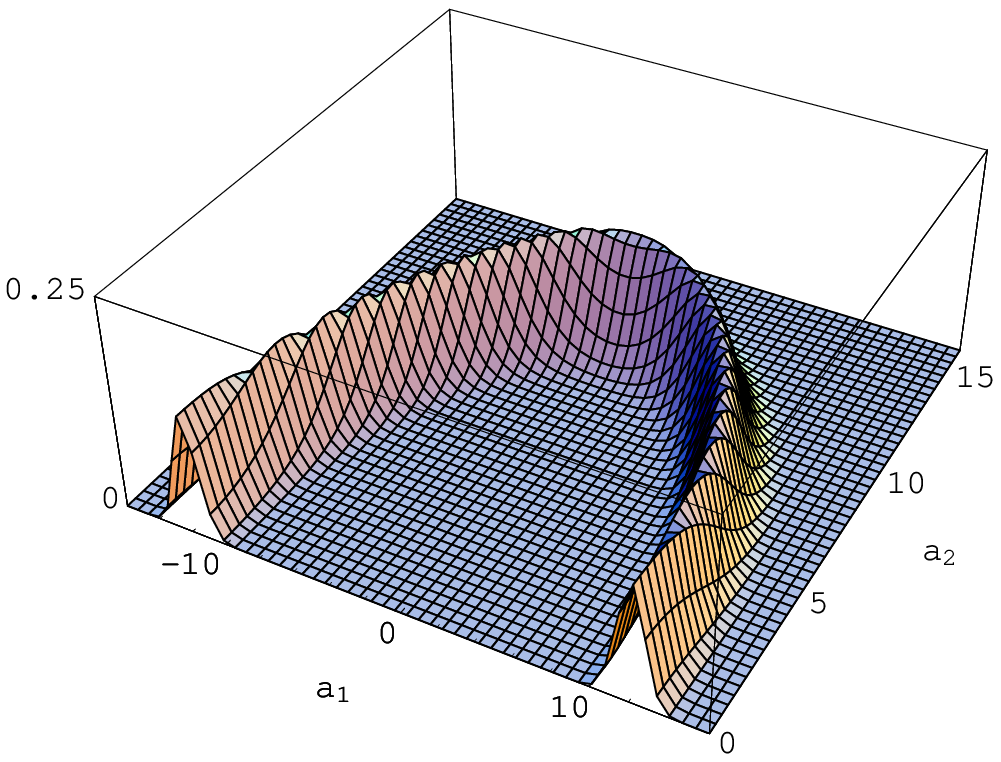,width=7cm}
 &\hspace{2cm}&
\epsfig{figure=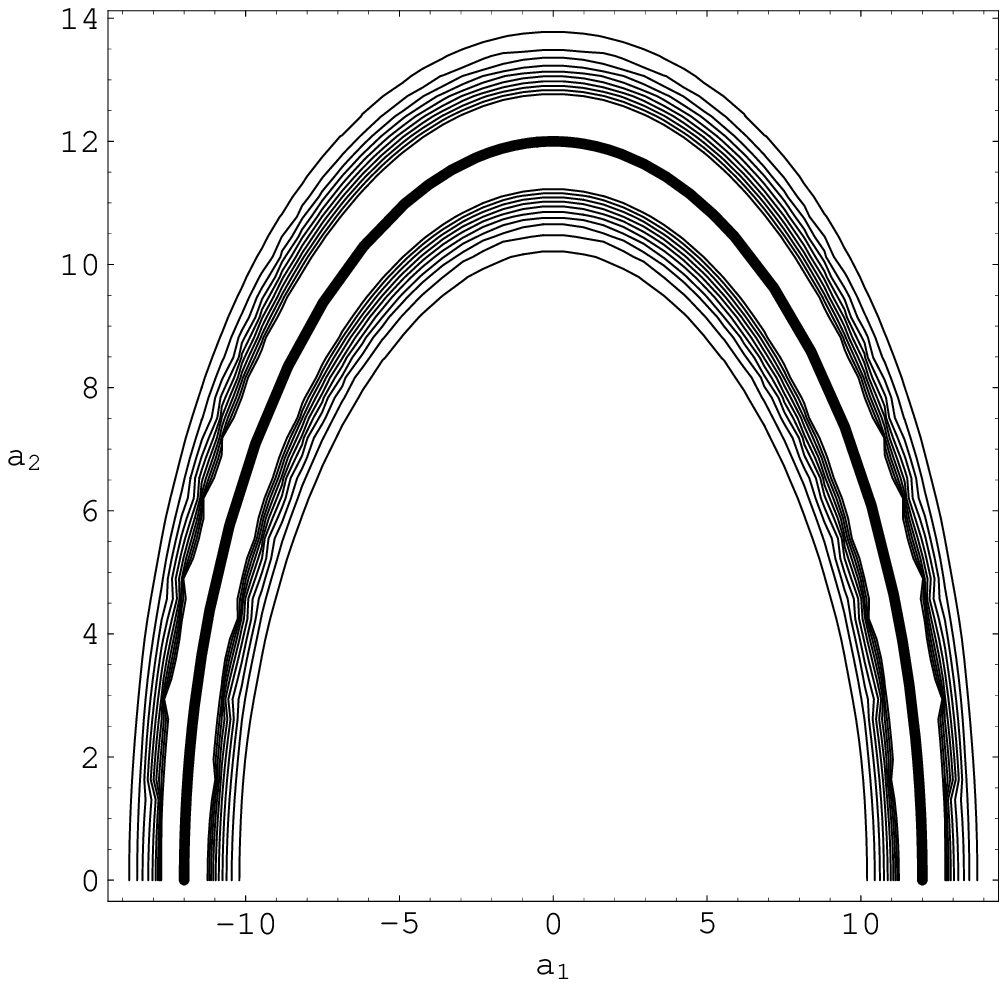,width=5.5cm}
\end{tabular}  }
\caption{\footnotesize Left, the square of the wave packet $|
\Psi(\alpha_1,\alpha_2)|^2$ for $|\chi_0|=12$ and $\theta_0 =0$,
$\omega_{+}=\omega_{-}=1$ with $n_{\mbox{max}}=130$ and right, the
contour plot of the same figure with the classical path
superimposed as the thick solid line.} \label{fig2}
\end{figure}
that is, the orthonormality and completeness of the basis
functions follow from those of the Hermite polynomials. A general
wave packet can now be defined as
\begin{equation}
\Psi (\alpha_{1},\alpha_{2})=\Sigma
_{n_{1},n_{2}}'A_{n_{1},n_{2}}\Phi_{n_{1},n_{2}}(\alpha_{1},\alpha_{2})
\label{4.8}
\end{equation}
where the prime on the sum indicates summing over all values of $
n_{1} $ and $ n_{2}$ satisfying the constraint (\ref{4.7}). The
coefficients $A_{n_{1},n_{2}}$ are given by \cite{gousheh}
\begin{eqnarray}
\frac{A_{n_{1},n_{2}}}{\sqrt{2^{n_2}n_2!}}=\left(\frac{\pi}{\omega_{-}}\right)^{1/4}
\frac{(n_{2}/2)! c_{n_{1}}}{(-1)^{n_{2}/2}n_{2}!} \label{4.9}
\end{eqnarray}
where
\begin{eqnarray}
c_{n}=e^{-1/4 |\chi_{0}|^{2}}\frac{\chi_{0}^{n}}{\sqrt{2^{n}n!}}.
\label{4.10}
\end{eqnarray}
The classical paths corresponding to these solutions are the
generalized Lissajous ellipsis which have the following parametric
representation
\begin{eqnarray}
\alpha_{1}(t)=\alpha_{1}(0) \cos(\omega_{+}t-\theta_{0}),
\hspace{0.5cm} \alpha_{2}(t)=\alpha_{2}(0) \sin(\omega_{-}t)
\label{4.11}
\end{eqnarray}
where the zero energy condition demands $
\omega_{+}\alpha_{1}(0)=\omega_{-}\alpha_{2}(0)$, and $ \theta_{0}
$ is an arbitrary phase factor. The classical-quantum
correspondence is established by
\begin{equation}
\chi_{0}=\sqrt{\frac{\omega_{+}}{\omega_{-}}} \alpha_{1}(0)
e^{i\theta_{0}}. \label{4.12}
\end{equation}
Figure \ref{fig2} shows the square of the wave packet
$|\Psi(\alpha_1,\alpha_2)|^2$ for equal frequencies
$\omega_+=\omega_-=1$ and also the corresponding classical path
superimposed on it, showing a good classical quantum
correspondence.  One can also choose the frequencies to be unequal
and study the the behavior of the resulting wave packet. An
extensive discussion for the construction of wave packets
resulting from the solutions of equation (\ref{4.3}) with both
equal and unequal frequencies can be found in \cite{gousheh}.

A final word on these solutions are in order. As the classical
solutions corresponding to these wave packets were shown to
undergo signature transition, the present quantum cosmological
solutions can also be said to have the same behavior in an
indirect way.

\section{conclusions}
In this paper we have considered the analytic expressions for a
class of degenerate metric solutions of the Einstein field
equations for a self-coupled scalar field in a $(4 +
d)$-dimensional cosmology with a FRW-type external metric. These
solutions depend on a free parameter $\delta$ whose limits are
specified by the present observations on the size of the universe.
They also predict signature transition from a Euclidean to a
Lorentzian domain. For  negative eigenvalues of $ \Sigma $, we
have shown that, within the context of the present work, signature
transition can provide a mechanism which would make the scale
factors to remain finite. This in turn causes the internal
dimension of the model to become stabilized. That is, signature
change selects those solutions which are stabilized. The beginning
of the Euclidean and ending of the Lorentzian domains are
characterized by a singular behavior of the scalar field. Although
our exact solutions have been obtained for a spatially flat
universe, we predict that the solutions corresponding to $k \neq 0
$ will generally show the same type of behavior, as has been shown
in \cite{ghafoori} for the much simpler case of a $4D$-FRW
cosmology. We have also studied the quantum cosmology of our model
for the special case when $mm'-n=0$ for which the corresponding WD
equation was found to have resulted from a Hamiltonian describing
an oscillator-ghost-oscillator system. These solutions show a good
classical-quantum correspondence in that the classical paths
coincide with the crest of the wave functions resulting from the
solution of the WD equation.

To conclude, the role of the potential in this and similar models
should be duly emphasized. It causes signature transition to
occur, stabilizes the internal degrees of freedom and is the vital
ingredient for inflation. In turn, signature transition provides a
mechanism through which stabilization of, and restriction on the
number of internal degrees of freedom can be achieved.
\vspace{1.0cm} \noindent \\
{\bf Acknowledgements}\vspace{0.3cm}\\
The authors would like to thanks S. S. Gousheh and H. Salehi for
useful discussions. SJ would like to thank the research council of
Shahid Beheshti University for financial support.

\end{document}